\documentclass[11pt,a4paper]{article}

\usepackage{graphicx}
\usepackage{color}
\usepackage{amsmath,amssymb}
\usepackage{bbold}
\usepackage{t1enc}
\usepackage{cite}

\usepackage{amsmath,amssymb}
\usepackage{amsfonts}
\usepackage{bbold}
\usepackage{latexsym} 
\usepackage{cancel}
\usepackage{graphicx, rotating}
\usepackage{epsfig}
\usepackage{color}
\usepackage[dvipsnames]{xcolor}
\usepackage{multirow}

\DeclareMathSymbol{\shortminus}{\mathbin}{AMSa}{"39}
\DeclareMathSymbol{\shm}{\mathbin}{AMSa}{"39}

\newcommand{\RE}{\operatorname{Re}}

\newcommand{\app}{a_{1 1}}
\newcommand{\azz}{a_{0 0}}
\newcommand{\amm}{a_{{\textstyle \scriptscriptstyle -}1 {\textstyle \scriptscriptstyle -}1}}
\DeclareMathOperator{\tr}{tr}
\newcommand{\smo}{{\textstyle \scriptscriptstyle -}1}
\newcommand{\smt}{{\textstyle \scriptscriptstyle -}2}

\begin{document}

\begin{center}
\begin{Large}
{\bf Tripartite entanglement in $H \to ZZ,WW$ decays}
\end{Large}

\vspace{0.5cm}
\renewcommand*{\thefootnote}{\fnsymbol{footnote}}
\setcounter{footnote}{0}
J.~A.~Aguilar-Saavedra \\[1mm]
\begin{small}
Instituto de F\'isica Te\'orica IFT-UAM/CSIC, c/Nicol\'as Cabrera 13--15, 28049 Madrid, Spain \\
\end{small}
\end{center}

\begin{abstract}
The decays of the Higgs boson produce a state in which the spins of the decay products and the orbital angular momentum ($L$) are highly entangled. We obtain the tripartite density operator, as well as reduced operators, for the decay into two weak bosons. For $H \to ZZ$ in the four-lepton final state we also estimate the statistical sensitivity at the Large Hadron Collider and future upgrades, using a binned method to reconstruct the density operators from distributions. With the expected Run 3 data, establishing genuine tripartite entanglement would be possible beyond the $5\sigma$ level. The violation of Bell inequalities involving the spins of the two weak bosons could also be established beyond the $5\sigma$ level. 
\end{abstract}

\section{Introduction}

High-energy physics experiments offer a new playground to test quantum entanglement at the energy frontier, with proposals to test the spin entanglement of pairs of top quarks~\cite{Afik:2020onf,Fabbrichesi:2021npl,Severi:2021cnj,Afik:2022kwm,Aguilar-Saavedra:2022uye,Afik:2022dgh,Dong:2023xiw,Han:2023fci}, muons~\cite{Aguilar-Saavedra:2023lwb}, $\tau$ leptons~\cite{Altakach:2022ywa}, weak bosons~\cite{Barr:2021zcp,Aguilar-Saavedra:2022wam,Aguilar-Saavedra:2022mpg,Fabbri:2023ncz,Ashby-Pickering:2022umy,Fabbrichesi:2023cev,Morales:2023gow}, and particles of different spin~\cite{Aguilar-Saavedra:2023hss,Aguilar-Saavedra:2024fig,Aguilar-Saavedra:2024hwd}. It is well known that, in addition to spin, particle production and decay involves orbital angular momentum (hereafter referred to as $L$). Therefore, in the aforementioned examples not only the spins of the two particles $S_1$, $S_2$ may be entangled, but they can also be entangled with $L$, as a consequence of total angular momentum conservation. An obvious step further is then to test the tripartite entanglement between $S_1$, $S_2$ and $L$ in these processes. 
The Hilbert space for $L$, in which the eigenstates $|lm\rangle$, with $l = 0,\dots\infty$ and $m= -l,\dots,l$ form an orthonormal basis, has infinite dimension in general. However, in decay processes the conservation of total angular momentum implies that only a finite-dimensional subspace $\mathcal{H}_L$ is relevant. For example, in Higgs boson decays to weak boson pairs $VV$, $V = W,Z$, we have $l \leq 2$, and $\mathcal{H}_L$ is 9-dimensional. Likewise, in $H \to \tau^+ \tau^-$, $\mathcal{H}_L$ has dimension four, with $l \leq 1$.

The Higgs boson is the only elementary scalar, and its decays produce highly-entangled states.
In this work we address the tripartite entanglement in $H \to VV$ decays, working in the Hilbert space $\mathcal{H}_L \otimes \mathcal{H}_{S_1} \otimes \mathcal{H}_{S_2}$. We use the method proposed in Ref.~\cite{Aguilar-Saavedra:2024vpd} to determine density operators involving $L$. In contrast to spin, the density operators involving $L$ cannot be directly measured from angular distributions. But they can be related, in a model-independent fashion, to measurable quantities. This is done by exploiting the complementarity between:
\begin{enumerate}
\item[(i)] Helicity amplitudes~\cite{Jacob:1959at}, in which the spin of the parent particle is quantised in some fixed axis $\hat z$, but for the decay products it is quantised in their direction of motion. These amplitudes depend on a small set of parameters that in principle can be measured in data.
\item[(ii)] Canonical amplitudes, in which the spin of all particles is quantised in a fixed direction $\hat z$. These amplitudes allow to build the multi-partite density operators involving $L$ and the spins of the decay products.
\end{enumerate}
The relation between these sets is model-independent, and arises from a simple change of basis for the spinors and polarisation vectors entering the amplitudes. It has been used in Ref.~\cite{Aguilar-Saavedra:2024vpd} to write down the full density operator for top quark decays $t \to W b$, which involves $L$ as well as the spins of the $W$ boson and $b$ quark. 
For $H \to VV$, the helicity amplitudes depend on three complex quantities. Once these quantities are measured in data, the full 81-dimensional density operator for $L$ and the spins of the two weak bosons can be determined. It turns out that $L$, $S_1$ and $S_2$ are genuinely entangled, that is, no bipartition of the $(L S_1 S_2)$ system is separable. Tracing over $L$ degrees of freedom, one obtains the density operator for the two spin degrees of freedom $S_1$ and $S_2$, which are highly entangled. Tracing over one of the spins $S_i$, $i=1,2$, the density operator for $L S_j$ ($j \neq i$) is found. We provide leading-order calculations for all these density operators in the Standard Model (SM) as well as methods to determine them from experimental data. We also provide estimates of the statistical sensitivity to establish the entanglement between the different parties in $H \to ZZ \to 4 \ell$, with $\ell = e,\mu$, at the Large Hadron Collider (LHC) and its future high-luminosity upgrade (HL-LHC).

\section{Higgs decay amplitudes}
\label{sec:2}

Let us consider a general two-body decay of a spin-$J$ particle with third spin component $M$, described in a coordinate system with a basis $\{ \hat x, \hat y, \hat z\}$ in its rest frame. In the helicity framework of Jacob and Wick~\cite{Jacob:1959at} the decay amplitudes  have the general form
\begin{equation}
A^h_{M \lambda_1 \lambda_2} (\theta,\phi) = a_{\lambda_1 \lambda_2} D_{M \lambda}^{J\,*} (\phi,\theta,0) \,,
\label{ec:aJW}
\end{equation}
where $\lambda_1$ and $\lambda_2$ are the helicities of the decay products, $\lambda = \lambda_1 - \lambda_2$, $a_{\lambda_1 \lambda_2}$ are independent of the angles and $D^j_{m'm}(\alpha,\beta,\gamma)$ are the Wigner functions
\begin{equation}
D^j_{mm'} \equiv \langle j m' | e^{-i \alpha J_3} e^{-i \alpha J_2} e^{-i \gamma J_3} | j m\! \rangle \,.
\end{equation}
The angles $(\theta ,\phi)$ correspond to the first decay product (with helicity $\lambda_1$). In the case of a scalar decay, $D^0_{00} = 1$, and therefore the helicities are equal. Furthermore, for a scalar decay into vector bosons, the helicity amplitudes are parameterised by three quantities  $\app$, $\azz$, $\amm$. The same parameters enter the canonical amplitudes. For an on-shell decay, these parameters are constant; however, for $H \to VV$ they depend on the invariant mass of the off-shell $V$ boson $m_{V^*}$, or equivalently on the modulus of the three-momenta in the Higgs rest frame, which we denote as $q$.

The canonical amplitudes $A^c_{s_1 s_2}$ (we omit the trivial index $M$) are found by changing the basis for the polarisation vectors. For a vector boson with momentum
\begin{equation}
p_V = (E_V, q \sin \theta \cos \phi, q \sin \theta \sin \phi, q \cos \theta) 
\end{equation}
the polarisation vectors in the helicity basis are
\begin{align}
& \varepsilon_h^{(+)} = - \frac{1}{\sqrt 2} (0,\cos \theta \cos \phi - i \sin \phi,\cos \theta \sin \phi + i \cos \phi,-\sin \theta) \,, \notag \\
& \varepsilon_h^{(0)} = \frac{1}{M_V} (q,E_V \sin \theta \cos \phi, E_V \sin \theta \sin \phi, E_V \cos \theta ) \,, \notag \\
& \varepsilon_h^{(-)} = \frac{1}{\sqrt 2} (0,\cos \theta \cos \phi + i \sin \phi,\cos \theta \sin \phi - i \cos \phi,-\sin \theta) \,.
\end{align}
On the other hand, the polarisation vectors in the fixed basis are
\begin{align}
& \varepsilon^{(+)} = \varepsilon_R^{(+)} - \frac{1}{\sqrt 2} \sin \theta e^{i \phi} \frac{q}{M_V} \left( 1, \frac{\vec p}{M_V+E_V} \right) \,, \notag \\
& \varepsilon^{(0)} = \varepsilon_R^{(0)} + \cos \theta \frac{q}{M_W} \left( 1, \frac{\vec p}{M_V+E_V} \right) \,, \notag \\
& \varepsilon^{(-)} = \varepsilon_R^{(-)} + \frac{1}{\sqrt 2} \sin \theta e^{-i \phi} \frac{q}{M_V} \left( 1, \frac{\vec p}{M_V+E_V} \right) \,,
\end{align}
where
\begin{align}
& \varepsilon_R^{(+)} = - \frac{1}{\sqrt 2} (0,1,i,0) \,, \notag \\
& \varepsilon_R^{(0)} = (0,0,0,1) \,, \notag \\
& \varepsilon_R^{(-)} = \frac{1}{\sqrt 2} (0,1,-i,0)
\label{ec:epsR}
\end{align}
are the $V$ rest-frame polarisation vectors. With this change of basis, the canonical amplitudes acquire a dependence on $(\theta,\phi)$ that is not present in the helicity amplitudes for a scalar decay. Expanding them in terms of spherical harmonics $Y_l^m$, and bearing in mind that $\langle \Omega | lm \rangle = Y_l^m(\Omega)$, with $\Omega = (\theta,\phi)$, one can identify the amplitudes into $L$ eigenstates $A_{s_1 s_2 ; l m}$. The non-zero ones are
\begin{align}
& A_{11;2-2} = A_{-1-1;22} = - \sqrt{\frac{2\pi}{15}} (\app + 2 \azz + \amm) \,, \notag \\
& A_{10;2-1}  = A_{01;2-1} =  A_{0-1;21} = A_{-10;21} = \sqrt{\frac{\pi}{15}} (\app + 2 \azz + \amm) \,, \notag \\
& A_{1-1;20} = A_{-11;20} = - \sqrt{\frac{2\pi}{45}} (\app + 2 \azz + \amm)  \,, \notag \\
& A_{00;20} = - \sqrt{\frac{4\pi}{45}} (\app + 2 \azz + \amm)  \,, \notag \\
& A_{10;1-1} = - A_{1-1;10} = - A_{01;1-1} = A_{0-1;11} = A_{-11;10} = - A_{-10;11} =  \sqrt{\frac{\pi}{3}} (\app - \amm) \,, \notag \\
& A_{1-1;00} =  - A_{00;00} = A_{-11;00} = - \sqrt{\frac{4\pi}{9}} (\app - \azz + \amm)  \,.
\end{align}
We finally note that, for an off-shell vector boson as in $H \to VV$, the propagator includes a scalar degree of freedom. Here we will be interested in decays into light charged leptons $\ell =e,\mu$ and neutrinos, and when coupled to massless external fermions the scalar component vanishes~\cite{Berge:2015jra}; therefore, we can safely consider the off-shell W as a spin-1 particle. This assumption is supported by the comparison of analytical and Monte Carlo calculations.

\section{Density operators for Higgs decay}
\label{sec:3}

\begin{figure}[htb]
\begin{center}
\includegraphics[height=6cm,clip=]{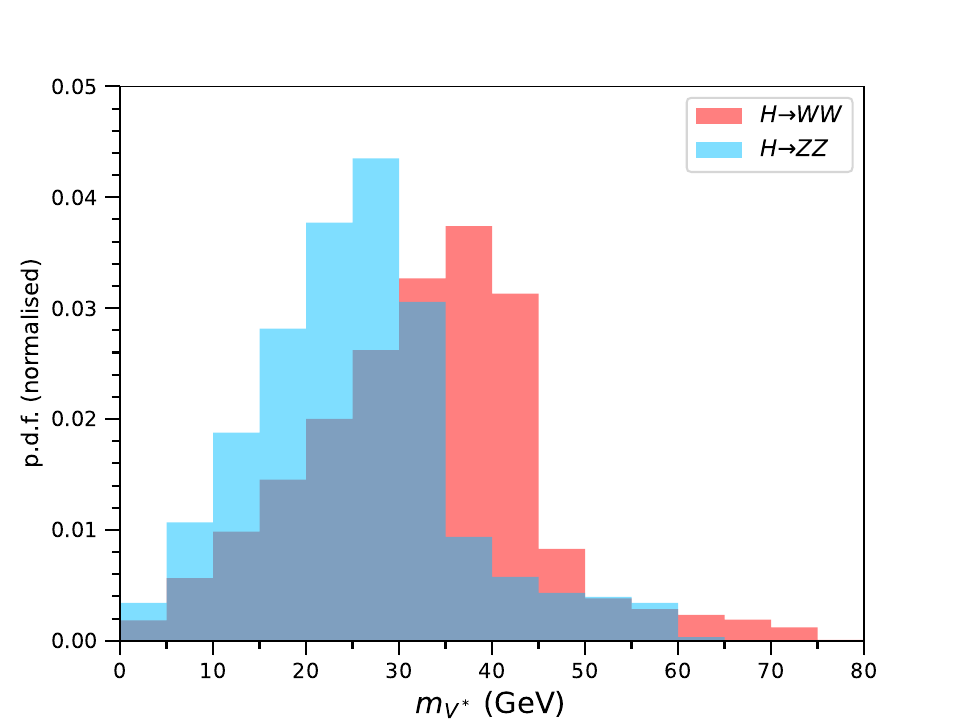} 
\caption{Invariant mass distribution of the off-shell boson in Higgs decays to $WW$ and $ZZ$.}
\label{fig:MV2}
\end{center}
\end{figure}

Angular momentum in the two-body decay $H \to VV$ is described by a density operator $\rho_{LS_1 S_2}$ acting on the Hilbert space $\mathcal{H}_L \otimes \mathcal{H}_{S_1} \otimes \mathcal{H}_{S_2}$. For fixed $m_{V^*}$ (or $q$), it can be calculated from the decay amplitudes found in the previous section as
\begin{equation}
(\rho_{LS_1 S_2})_{s_1 s_2; l m}^{s_1' s_2'; l' m'} = A_{s_1 s_2;lm} A_{s_1' s_2';l'm'}^* \,.
\label{ec:rho3}
\end{equation}
Notice that $\rho_{LS_1 S_2}$ as calculated from this equation always corresponds to a pure state.
As mentioned, $\app$, $\azz$, $\amm$ depend on $m_{V^*}$. The kinematical distribution of this quantity, calculated with {\scshape Madgraph}~\cite{Alwall:2014hca} at the leading order (LO), is presented in Fig.~\ref{fig:MV2} for $H \to WW$ and $H \to ZZ$.
In practice, Eq.~(\ref{ec:rho3}) can be used to calculate the density operator within some interval of $m_{V^*}$. For example, in $H \to ZZ$ with $m_{Z^*} \in [18,32]$ GeV (which amounts to 55\% of the events) the exact density operator calculated with Monte Carlo describes a pure state up to corrections of the order of $0.5\%$. Likewise, for $H \to WW$ with $m_{W^*} \in [30,44]$ GeV (49\% of the events) the density operator corresponds to a pure state up to corrections of 1\%. Therefore, using (\ref{ec:rho3}) is quite a good approximation even for bin widths of the order of 10 GeV.  

In order to obtain $\rho_{LS_1 S_2}$ for the full decay phase space we use Monte Carlo calculations of $gg \to H \to ZZ \to e^+ e^- \mu^+ \mu^-$ and $gg \to H \to W^+W^- \to \ell^+ \nu \ell^- \nu$ with {\scshape Madgraph} at LO, using $7 \times 10^6$ and $10^7$ events, respectively. 
In $H \to ZZ$, we label the boson with largest invariant mass as $V_1$, and in $H \to W^+ W^-$ we select $V_1 = W^+$. (In any case, the predictions are symmetric under interchange $1 \leftrightarrow 2$.)
We divide the $m_{V^*}$ range in 2 GeV intervals and, within each bin `$k$', the values of $\app$, $\azz$ and $\amm$ are extracted from Monte Carlo pseudo-data (see section~\ref{sec:5} for details) using parton-level information.\footnote{As a cross-check, we do the same in 5 GeV intervals. The numerical results are the same, with differences in the range $10^{-4}-10^{-3}$.}
The density operator $\rho_{LS_1 S_2}^{(k)}$ for that bin is calculated using (\ref{ec:rho3}), and the theoretical prediction for $\rho_{LS_1 S_2}$ in the full $m_{V^*}$ range is obtained by summing the operators $\rho_{LS_1 S_2}^{(k)}$ in the different bins, with the appropriate weight. This density operator no longer describes a pure state, but it is quite close. For $H \to ZZ$, the principal eigenvector has eigenvalue 0.966 (where unity would correspond to a pure state), and in $H \to WW$ the principal eigenvector has eigenvalue 0.963.

Tracing over the Hilbert space of any of the subsystems $\mathcal{H}_L$, $\mathcal{H}_{S_1}$, or $\mathcal{H}_{S_2}$, we obtain the reduced density operators for the other two, respectively $\rho_{S_1 S_2}$, $\rho_{L S_2}$, and $\rho_{L S_1}$. These operators correspond to the marginalisation over the corresponding degree of freedom. Tracing over two of the subsystems one obtains the density operator for a single subsystem. For the spin degrees of freedom, and since the scalar decay does not have any preferred direction, the operators describe a completely unpolarised state,
\begin{equation}
\rho_{S_1} = \rho_{S_2} = \frac{1}{3} \left( \! \begin{array}{ccc} 1 & 0 & 0 \\ 0 & 1 & 0 \\ 0 & 0 & 1 \end{array} \! \right) \,,
\label{ec:rhoS}
\end{equation}
by construction.
For $L$, the density operator is diagonal, and for fixed $m_{V^*}$ it has entries
\begin{align}
& (\rho_L)_{2m}^{2m} = \frac{1}{30} \frac{1}{\mathcal{N}} |\app + 2 \azz + \amm|^2 \,, \notag \\
& (\rho_L)_{1m}^{1m} = \frac{1}{6} \frac{1}{\mathcal{N}} |\app - \amm|^2 \,, \notag \\
& (\rho_L)_{00}^{00} =  \frac{1}{3} \frac{1}{\mathcal{N}} |\app - \azz + \amm|^2 \,,
\end{align}
with
\begin{equation}
\mathcal{N} = |\app|^2 + |\azz|^2 + |\amm|^2 \,.
\end{equation}
For the full phase space, the Monte Carlo calculation yields $(\rho_L)_{2m}^{2m} = 0.021$, $(\rho_L)_{1m}^{1m} = 0$, $(\rho_L)_{00}^{00} = 0.895$ for both $H \to ZZ$ and $H \to WW$. It is worthwhile pointing out that, even if the Higgs decay is isotropic, there are $l=2$ contributions (as well as $l=1$ contributions if CP is broken in the decay). Note that
\begin{equation}
\sum_{m=-2}^2 |Y_2^m (\theta,\phi)|^2 = \frac{5}{4\pi} \,, \quad
\sum_{m=-1}^1 |Y_1^m (\theta,\phi)|^2 = \frac{3}{4\pi} \,.
\label{ec:Ysum}
\end{equation}
This cancellation of the angular dependence makes apparent the need to determine $\rho_L$ indirectly from the measurement of $\app$, $\azz$ and $\amm$, because one cannot directly access it through distributions~\cite{Aguilar-Saavedra:2024vpd}.

The entanglement between one of the subsystems ($L$, $S_1$ or $S_2$) and the rest can be tested by taking the partial transpose of $\rho_{LS_1 S_2}$ over the corresponding space $\mathcal{H}_L$, $\mathcal{H}_{S_1}$, or $\mathcal{H}_{S_2}$. For a bipartite system $AB$ (such as $\mathcal{H}_L$ versus $\mathcal{H}_{S_1} \otimes \mathcal{H}_{S_2}$) the entanglement can be characterised by
the Peres-Horodecki~\cite{Peres:1996dw,Horodecki:1997vt} criterion. Since the positivity of the partial transpose over any subsystem, say $\rho^{T_B}$, is a necessary condition for separability, a non-positive $\rho^{T_B}$ is a sufficient condition for entanglement. Furthermore, the amount of entanglement can be quantified by the negativity of $\rho^{T_B}$~\cite{Plenio:2007zz},
\begin{equation}
N(\rho) = \frac{\| \rho^{T_B} \| - 1}{2} \,,
\label{ec:Nrho}
\end{equation}
where $\|X\| = \tr \sqrt{XX^\dagger} = \sum_i \sqrt{\lambda_i}$, where $\lambda_i$ are the (positive) eigenvalues of the matrix $XX^\dagger$. Equivalently, $N(\rho)$ equals the sum of the negative eigenvalues of $\rho^{T_B}$. (The result is the same when taking the partial transpose on subsystem $A$.) In the separable case $N(\rho) = 0$.
For pure states the generalised concurrence~\cite{PhysRevA.64.042315} can also be used as entanglement measure. For a bipartite system $AB$, it is defined as
\begin{equation}
C^2 = 2 (1 - \operatorname{tr} \rho_A^2) \,,
\end{equation}
with $\rho_A$ the reduced density operator obtained by trace over the $B$ degrees of freedom. The result is the same when tracing over $\mathcal{H}_A$, and in the separable case $C^2 = 0$.

These tests, performed for the three possible bipartitions, give sufficient conditions for genuine tripartite entanglement, that is, that the state is not separable under any bipartition of $\mathcal{H}_L \otimes \mathcal{H}_{S_1} \otimes \mathcal{H}_{S_2}$. The entanglement measures between the different bipartitions are collected in Table~\ref{tab:trip}, for the full decay phase space as well as on selected bins of $m_{V^*}$, in which the Higgs decay produces a pure state to an excellent approximation.\footnote{A global measure of the tripartite entanglement for unequal systems of these dimensions ($9 \times 3 \times 3$) is not currently available in the literature.}

\begin{table}[h]
\begin{center}
\begin{tabular}{lcccc}
& $L$-$(S_1 S_2)$ & $S_1$-$(L S_2)$ & $S_2$-$(L S_1)$ \\
                               
$H \to ZZ$  inclusive & $N = 0.757$ & $N = 0.998$ & $N = 0.998$ \\[1mm]
\multirow{2}{*}{$H \to ZZ$, $m_{Z^*} \in [25,30]$ GeV}   & $N = 0.421$ & $N = 1$ & $N = 1$ \\
                                                                                         & $C^2 = 0.108$ & $C^2 = 4/3$ & $C^2 = 4/3$ \\[1mm]
$H \to WW$ inclusive & $N = 0.746$ & $N = 0.998$ & $N = 0.998$ \\[1mm]
\multirow{2}{*}{$H \to WW$, $m_{W^*} \in [35,40]$ GeV} & $N = 0.376$ & $N = 1$ & $N = 1$ \\
                                                                                          & $C^2 = 0.088$ & $C^2 = 4/3$ & $C^2 = 4/3$ 

\end{tabular}
\caption{Entanglement measures for the different bipartitions of $\mathcal{H}_L \otimes \mathcal{H}_{S_1} \otimes \mathcal{H}_{S_2}$.}
\label{tab:trip}
\end{center}
\end{table}

For pure states the entanglement between one spin $S_{i}$ and the rest of the system is maximal. This is easily understood because the Higgs decay is isotropic: tracing over the $(L S_j)$ subsystem ($j \neq i$) yields the reduced operator (\ref{ec:rhoS}), which corresponds to an unpolarised state. Consequently, for the $S_i$-$(L S_j)$ entanglement the concurrence $C^2$ takes its maximal value for systems of the dimensionality ($9 \times 3 \times 3$) under consideration. The tripartite operators transposed in the spaces $\mathcal{H}_{S_1}$ or $\mathcal{H}_{S_2}$ have three eigenvalues $-1/3$, so that for $S_i$-$(L S_j)$ entanglement one has $N = 1$. Both $C^2 = 4/3$ and $N = 1$ are guaranteed by construction when one writes the $\rho_{LS_1 S_2}$ operator from Eq.~(\ref{ec:rho3}).
For the full phase space, in which $\rho_{LS_1 S_2}$ is a weighted sum of pure states $\rho_{LS_1 S_2}^{(k)}$, $N$ is still quite close to unity. A similar property does not hold for $L$-$(S_1 S_2)$ entanglement. As it follows from Eqs.~(\ref{ec:Ysum}) and the related discussion, there are many (infinite) possibilities for $\rho_L$ consistent with a Higgs isotropic decay.

One can also investigate the entanglement between a pair of subsystems, when the third one is marginalised. The eigenvalue $\lambda_\text{max}$ of the principal eigenvector is also of interest, in order to assess to which extent the reduced density operators correspond to a pure state.  Both are presented in Table~\ref{tab:bip}. The $L$-$S_i$ subsystems are weakly entangled, and in a very mixed state with $\lambda_\text{max} \leq 1/3$. This is in agreement with the fact that the entanglement between one spin and the rest of the system is maximal, as discussed above. On the other hand, the entanglement between the two spins is quite large, and their state, even after tracing over $L$ degrees of freedom, is relatively close to a pure state.

\begin{table}[h]
\begin{center}
\begin{tabular}{lcccc}
& $L$-$S_1$ & $L$-$S_2$ & $S_1$-$S_2$ \\
\multirow{2}{*}{$H \to ZZ$ inclusive}   & $\lambda_\text{max} = 0.322$ & $\lambda_\text{max} = 0.322$ & $\lambda_\text{max} = 0.896$ \\ 
                                                            & $N = 0.105$ & $N = 0.105$ & $N = 0.843$ \\[1mm]
\multirow{2}{*}{$H \to ZZ$, $M_{Z^*} \in [25,30]$ GeV}  & $\lambda_\text{max} = 1/3$ & $\lambda_\text{max} = 1/3$ & $\lambda_\text{max} = 0.972$ \\ 
                                                                                        & $N = 0.038$ & $N = 0.038$ & $N = 0.959$ \\[1mm]
\multirow{2}{*}{$H \to WW$ inclusive}  & $\lambda_\text{max} = 0.321$ & $\lambda_\text{max} = 0.321$ & $\lambda_\text{max} = 0.895$ \\ 
                                                             & $N = 0.102$ & $N = 0.102$ & $N = 0.843$ \\[1mm]
\multirow{2}{*}{$H \to WW$, $M_{W^*} \in [35,40]$ GeV} & $\lambda_\text{max} = 1/3$ & $\lambda_\text{max} = 1/3$ & $\lambda_\text{max} = 0.977$ \\ 
                                                                                          & $N = 0.031$ & $N = 0.031$ & $N = 0.966$
\end{tabular}
\caption{Eigenvalue of the principal eigenvector and negativity for the different pairs of systems when the third one is traced out.}
\label{tab:bip}
\end{center}
\end{table}

\section{Spin entanglement and Bell inequalities}
\label{sec:4}

The spin entanglement between the two weak bosons has already been addressed for $H \to WW$~\cite{Barr:2021zcp,Aguilar-Saavedra:2022mpg,Ashby-Pickering:2022umy,Fabbri:2023ncz,Fabbrichesi:2023cev} and $H \to ZZ$~\cite{Aguilar-Saavedra:2022wam,Ashby-Pickering:2022umy,Fabbrichesi:2023cev} in the helicity basis $\{ \hat r, \hat n, \hat k\}$ where the quantisation axis $\hat k$ is taken in the flight direction of $V_1$, $\hat k = (\theta,\phi)$. The density operator is obtained after integration over all decay phase space with this `moving' coordinate system. The angular integration is trivial  in this case because the Higgs decay is isotropic. On the other hand, the operator $\rho_{S_1 S_2}$ obtained in the previous section parameterises the spin state of the $VV$ pair with a fixed basis $\{ \hat x, \hat y, \hat z\}$ and integrated over all decay phase space --- the angular integration is not trivial in this case because we fix a preferred direction. Both descriptions are not equivalent. In the former case, for $H \to ZZ$ (and similarly for $H \to WW$) the density operator has an eigenvector
\begin{equation}
|\psi\rangle = 0.444 \left[ |1 \shm\! 1\rangle_{\hat k} +  |\shm\! 1 \, 1\rangle_{\hat k} \right]  - 0.777 \, |0\,0\rangle_{\hat k} \,,
\label{ec:PsiZZ}
\end{equation}
with eigenvalue 0.970. In the latter, the density operator has as principal eigenvector the spin singlet
\begin{equation}
|\psi\rangle = \frac{1}{\sqrt 3} \left[ |1 \shm\! 1\rangle - |0 \, 0\rangle + |\shm\! 1 \, 1\rangle \right]
\end{equation}
with eigenvalue 0.896. The numerical value of the entanglement measure $N$ is quite similar in both cases, however.

In this section we address spin entanglement and possible violation of Bell inequalities using the reduced density operator in the canonical basis as obtained in the previous section. We parameterise $\rho_{S_1 S_2}$ in terms of irreducible tensor operators 
$T^l_m$, with $l=1,2$ and $-l \leq m \leq l$, acting on the three-dimensional spin space for each boson~\cite{Aguilar-Saavedra:2022wam}. For convenience we normalise $T^l_m$ such that $\tr\left[T^l_m \left(T^l_m\right)^{\dagger}\right] = 3 $, where $\left(T^l_m\right)^{\dagger}=(-1)^m \, T^l_m $. Specifically, the operators are defined as
\begin{align}
& T^1_{\pm 1} = \mp \frac{\sqrt{3}}{2}  (J_1 \pm i J_2) \,,\quad T^1_0 = \sqrt{\frac{3}{2}} J_3 \,, \notag \\
& T^2_{\pm 2} = \frac{2}{\sqrt{3}} (T_{\pm 1}^1)^2 \,, \quad
    T^2_{\pm 1} = \sqrt{\frac{2}{3}} \left[T_{\pm 1}^1 T_{0}^1 + T_{0}^1 T_{\pm 1}^1 \right] \,, \notag \\
& T^2_0 = \frac{\sqrt{2}}{3} \left[T_1^1 T_{-1}^1 + T_{-1}^1 T_{1}^1 + 2 (T_{0}^1)^2 \right] \,,
\end{align}
with $J_i$ the usual spin operators in the Cartesian basis. In terms of these, the spin density operator reads~\cite{Aguilar-Saavedra:2022wam}
\begin{eqnarray}
\rho_{S_1 S_2} & = & \frac{1}{9}\left[
\mathbb{1}_3 \otimes \mathbb{1}_3 + A^1_{lm} \, T^l_{m} \otimes \mathbb{1}_3 + A^2_{lm} \, \mathbb{1}_3 \otimes T^l_{m}  + C_{l_1 m_1 l_2 m_2} \, T^{l_1}_{m_1} \otimes T^{l_2}_{m_2}
\right] \,,
\label{ec:rhoAC}
\end{eqnarray}
with a sum over all indices. Because $\rho_{S_1 S_2}$ is Hermitian, the coefficients satisfy the relations
\begin{align}
& (A^{1,2}_{lm})^* = (-1)^m \, A^{1,2}_{l \; -m} \,, \notag \\
& (C_{l_1 m_1 l_2 m_2})^* = (-1)^{m_1+m_2} \, C_{l_1 \; -m_1 l_2 \; -m_2} \,.
\end{align}
For fixed $m_{V^*}$, it is found that the only non-zero coefficients are\footnote{These relations different from (and not to be confused with) the ones found for the helicity basis~\cite{Aguilar-Saavedra:2022mpg}.}
\begin{eqnarray}
C_{1010} & = & - C_{111\smo} = - C_{1\smo11} \equiv C_1 \,, \notag \\
C_{2020} & = & C_{222\smt} = C_{2\smt 22} = - C_{212\smo} = - C_{2\smo 21} \equiv C_2 \,,
\label{ec:Crel}
\end{eqnarray}
with
\begin{eqnarray}
C_1 & = & \frac{1}{2} \frac{1}{\mathcal{N}} \left\{ - |\app|^2 - |\amm|^2 + 2 \RE \left[ (\app + \amm) \azz^* \right]  \right\}  \,, \notag \\
C_2 & = & \frac{1}{2} \frac{1}{\mathcal{N}} \left\{ |\app|^2 + 4 |\azz|^2 + |\amm|^2 - 6 \RE \left[ (\app + \amm) \azz^* \right] \right. \notag \\
& & \left. + 12 \RE \left[\app \amm^* \right] \right\} \,.
\label{ec:C1C2}
\end{eqnarray}
For the full decay phase space the relations (\ref{ec:C1C2}) with helicity amplitudes do not hold, but the relations between coefficients (\ref{ec:Crel}) still do because the density operator depends linearly on them. Therefore, the full phase-space reduced operator $\rho_{S_1 S_2}$ can be written as the expansion (\ref{ec:rhoAC}) in terms of two independent coefficients $C_1$ and $C_2$. The eigenvalues of the partial transpose $\rho_{S_1 S_2}^{T_B}$ are
\begin{eqnarray}
\lambda_5 & = & \frac{1}{18} (2 - 3 C_1 + C_2) \quad \text{(quintuple)}  \,, \notag \\
\lambda_3 & = & \frac{1}{18} (2 + 3 C_1 - 5 C_2) \quad \text{(triple)} \,, \notag \\ 
\lambda_1 & =  & \frac{1}{9} (2 + 3 C_1 + 5 C_2) \,.
\end{eqnarray}
By the Peres-Horodecki criterion, finding any of these eigenvalues negative is a sufficient condition for spin entanglement. 
Given the SM values $C_1 = -0.844$, $C_2 = 0.906$ (both for $ZZ$ and $WW$), only $\lambda_3$ is expected to be negative, $\lambda_3 = -0.281$, and a useful entanglement test that is equivalent to the negativity, $N = -3 \lambda_3$.

Concerning Bell-like inequalities, for a Hilbert space $\mathcal{H}_A \otimes \mathcal{H}_B$ with both subsystems $A$, $B$ having dimension 3, a powerful test is provided by the Collins-Gisin-Linden-Massar-Popescu (CGLMP) inequality~\cite{Collins:2002sun}: for observables $A_1$, $A_2$ in $\mathcal{H}_A$, and observables $B_1$, $B_2$ in $\mathcal{H}_B$, we have
\begin{eqnarray}
I_3 &=& P(A_1=B_1) + P(B_1=A_2+1) + P(A_2=B_2) + P(B_2=A_1) - P(A_1=B_1-1)  \notag \\
& & - P(B_1=A_2) - P(A_2=B_2-1) - P(B_2=A_1-1) \leq 2
\label{CGLMP}
\end{eqnarray}
in any local realistic theory. Here, $P(B_i = A_j + a)$ is the probability that the measurements of $B_i$ gives the same result as the measurement of $A_j$ plus $a$, modulo 3. This inequality can be conveniently written in terms of Bell operators $\mathcal{O}_\text{Bell}$, such that 
\begin{equation}
I_3 \equiv \langle \mathcal{O}_\text{Bell} \rangle = \tr [\mathcal{O}_\text{Bell} \, \rho_{S_1 S_2}] \leq 2
\label{ec:cglmp}
\end{equation}
in a local realistic theory. A choice of Bell operator that is optimal for the maximally-entangled spin-singlet state is~\cite{Acin:2002zz}
\begin{equation}
\mathcal{O}_\text{Bell} = \frac{4}{3 \sqrt 3} (T^1_1 \otimes T^1_{-1} + T^1_{-1} \otimes T^1_1 )  +
 \frac{2}{3} (T^2_2 \otimes T^2_{-2} + T^2_{-2} \otimes T^2_2 )  \,.
\end{equation}
However, the $VV$ pair is not produced in a pure spin-singlet state. An operator that gives a larger value for $I_3$ in the helicity basis was found in Ref.~\cite{Aguilar-Saavedra:2022wam},
\begin{eqnarray}
\mathcal{O}'_\text{Bell} & = & -\frac{4}{3 \sqrt 3} T^1_0 \otimes T^1_0 + \frac{1}{2} T^2_0 \otimes T^2_0
+ \frac{2}{3 \sqrt 3} (T^1_1 \otimes T^1_{-1} + T^1_{-1} \otimes T^1_1 ) \notag \\
& & - \frac{1}{3} (T^2_1 \otimes T^2_{-1} + T^2_{-1} \otimes T^2_1 )
+ \frac{1}{12} (T^2_2 \otimes T^2_{-2} + T^2_{-2} \otimes T^2_2 )  \,.
\end{eqnarray}
In the canonical basis, and given the relations (\ref{ec:Crel}) between coefficients, both operators are equivalent. 
We compare in Table~\ref{tab:Obell} the values obtained for $I_3$ in the helicity basis (as used in previous works) and in the canonical basis, using both operators. 

\begin{table}[h]
\begin{center}
\begin{tabular}{lcccc}
& \multicolumn{2}{c}{$H \to ZZ$} & \multicolumn{2}{c}{$H \to WW$} \\
& $\mathcal{O}_\text{Bell}$ & $\mathcal{O}'_\text{Bell}$ & $\mathcal{O}_\text{Bell}$ & $\mathcal{O}'_\text{Bell}$ \\ 
Helicity & 2.327 & 2.691 & 2.270 & 2.629 \\
Canonical & \multicolumn{2}{c}{2.507} & \multicolumn{2}{c}{2.506} 
\end{tabular}
\caption{Values of the quantity $I_3$ signaling violation of the CGLMP inequalities.}
\label{tab:Obell}
\end{center}
\end{table}

\section{Parameter determination}
\label{sec:5}

A model-independent measurement of $\app$, $\azz$ and $\amm$ and their relative phases is possible from angular distributions. 
For the decay $V_1 V_2 \to f_1 f_1' f_2 f_2'$ we label as $(\theta_{1,2},\phi_{1,2})$ the polar and azimuthal angles of $f_{1,2}$ in the $V_{1,2}$ rest frame, with respect to some coordinate system to be specified later. Then, corresponding to a density operator of the form (\ref{ec:rhoAC}), the four-dimensional angular distribution is~\cite{Aguilar-Saavedra:2022wam,Aguilar-Saavedra:2022mpg}
\begin{eqnarray}
\frac{1}{\sigma}\frac{d\sigma}{d\Omega_1d\Omega_2} & = & \frac{1}{(4\pi)^2}\left[ 1 +A_{lm}^1 B_l Y_l^m(\theta_1, \phi_1) + A_{lm}^2 B_l Y_l^m(\theta_2, \phi_2)   \right. \notag \\
& & \left. + C_{l_1 m_1 l_2 m_2} B_{l_1}B_{l_2} Y_{l_1}^{m_1}(\theta_1, \phi_1)Y_{l_2}^{m_2}(\theta_2, \phi_2)  \right] \,,
\label{ec:dist4D}
\end{eqnarray}
with $B_1$, $B_2$ constants. For $H \to ZZ$, and taking $f_{1,2}$ as the negative leptons, one has
\begin{eqnarray}
B_1=-\sqrt{2\pi} \eta_\ell\ , \ \ \ B_2=\sqrt{\frac{2\pi}{5}}
\label{ec:BL}
\end{eqnarray}
with
\begin{equation}
\eta_\ell = \frac{1-4 s_W^2}{1-4 s_W^2 + 8 s_W^4} \simeq 0.13 \,,
\label{ec:etal}
\end{equation}
$s_W$ being the sine of the weak mixing angle. For $H \to WW$, $B_{1,2}$ are as in (\ref{ec:BL}) setting $\eta_\ell = 1$ for $\ell^-$ and $\eta_\ell = -1$ for $\ell^+$. 

As we have remarked in the introduction, the extraction of the parameters $\app$, $\azz$ and $\amm$ from data (or pseudo-data) is done by using the helicity basis $\{ \hat r, \hat n, \hat k\}$, defined as follows:
\begin{itemize}
\item The $\hat k$ axis is taken in the direction of the $V_1$ three-momentum in the Higgs boson rest frame.
\item The  $\hat r$ axis is in the production plane and defined as $\hat r = \mathrm{sign}(\cos \theta) (\hat p_p - \cos \theta \hat k)/\sin \theta$, with $\hat p_p = (0,0,1)$ the direction of one proton in the laboratory frame, $\cos \theta = \hat k \cdot \hat p_p$. The definition for $\hat r$  is the same if we use the direction of the other proton $- \hat p_p$. 
\item The $\hat n$ axis is taken such that $\hat n = \hat k \times \hat r$, orthogonal to the production plane.
\end{itemize}
This reference system is used to measure the angles $(\theta_{1,2},\phi_{1,2})$. Implicitly, this requires the reconstruction of the rest frames of the decaying bosons. This is straightforward for $H \to ZZ \to 4\ell$, whereas for $H \to WW \to \ell \nu \ell \nu$ the system is underconstrained and the kinematics cannot be uniquely determined. Promising attempts have been made in this direction using machine-learning techniques~\cite{Grossi:2020orx}. Alternatively, one can consider the semi-leptonic decay mode $H \to WW \to \ell \nu q \bar q'$~\cite{Fabbri:2023ncz}, in which the full reconstruction is possible and the discrimination between jets originating from up- and down-type quarks is achieved with charm tagging.

In the remainder of this section we show how the parameters $\app$, $\azz$, $\amm$ can be extracted from data. We will present a model-independent extraction, followed by a method that assumes CP conservation in the $H \to VV$ decay.

\subsection{Model-independent parameter determination}

In the helicity basis, and considering fixed $m_{V^*}$, the non-zero coefficients in the expansion (\ref{ec:rhoAC}) are given by~\cite{Aguilar-Saavedra:2022mpg}\footnote{These relations are different from (and not to be confused with) the ones presented for the canonical basis in section~\ref{sec:4}.}
\begin{align}
& A_{20}^1 = A_{20}^2 = \frac{1}{\sqrt 2} \frac{1}{\mathcal{N}} \left[ 
|a_{1 1}|^2 + |a_{\smo \smo}|^2 - 2 |a_{00}|^2 \right] \,, \notag \\
& C_{1010} = - \frac{3}{2} \frac{1}{\mathcal{N}} \left[ 
|a_{1 1}|^2 + |a_{\smo \smo}|^2 \right] \,, \notag \\
& C_{2020} = \frac{1}{2} \frac{1}{\mathcal{N}} \left[ 
|a_{1 1}|^2 + |a_{\smo \smo}|^2 + 4 |a_{00}|^2 \right] \,, \notag \\
& C_{2 2 2 \smt} = C_{2 \smt 22}^* = 3 \frac{1}{\mathcal{N}} \; a_{11} a_{\smo \smo}^* \,, \notag \\
& C_{111 \smo} = - C_{212 \smo} = C_{1 \smo 11}^* = - C_{2 \smo 21}^*  = - \frac{3}{2} \frac{1}{\mathcal{N}} \left[ 
a_{11} a_{00}^* + a_{00} a_{\smo \smo}^* \right] 
\label{ec:C}
\end{align}
and, when CP is broken,
\begin{align}
& A_{10}^1 = - A_{10}^2 = \sqrt{\frac{3}{2}} \frac{1}{\mathcal{N}} \left[ 
|a_{1 1}|^2 - |a_{\smo \smo}|^2 \right] \,, \notag \\
& C_{1020} = - C_{2010} = \frac{\sqrt 3}{2}  \frac{1}{\mathcal{N}} \left[ 
|a_{1 1}|^2 - |a_{\smo \smo}|^2 \right] \,, \notag \\
& C_{1 \smo 2 1} = - C_{2 \smo 11} = C_{112 \smo}^* = - C_{211 \smo}^*  = \frac{3}{2} \frac{1}{\mathcal{N}} \left[ 
a_{00} a_{11}^* - a_{\smo \smo} a_{00}^* \right] \,.
\end{align}
In the following we set $\mathcal{N} = 1$ for simplicity, as the global normalisation is irrelevant. By using the relation
 $|\app|^2 + |\azz|^2 + |\amm|^2 = 1$, one can combine the measurements of $A_{20}^1$ and $A_{20}^2$ to find $|\azz|$, and the measurements of $A_{10}^1$ and $A_{10}^2$ to determine $|\app|$ and $|\amm|$. Alternatively, one can use the $\cos \theta_{1,2}$ distributions. By integrating Eq.~(\ref{ec:dist4D}) and using the above relations, one obtains
\begin{eqnarray}
\frac{1}{\sigma} \frac{d\sigma}{d\cos \theta_1} & = & \frac{3}{8} |\app|^2 (1 - 2 \eta_\ell \cos \theta_1 + \cos^2 \theta_1) 
+ \frac{3}{4} |\azz|^2 \sin^2 \theta_1 \notag \\
& &  + \frac{3}{8} |\amm|^2  (1 + 2 \eta_\ell \cos \theta_1 + \cos^2 \theta_1) \,,
\end{eqnarray}
and likewise for $\theta_2$. For $W$ boson decays this distribution is well known~\cite{Kane:1991bg}. A fit to the distribution with the constraint $|\app|^2 + |\azz|^2 + |\amm|^2 = 1$ provides the three moduli. Again, the distributions for $\theta_1$ and $\theta_2$ can be used to improve the determination. The relative phase between $\app$ and $\amm$ is found from the measurement of $C_{222\smt}$. The relative phase between $\app$ and $\azz$ can be obtained by measuring for example $C_{212\smo}$ and $C_{2\smo 11}$.

\subsection{Determination in the CP-conserving case}

Within the SM, CP is conserved in $H \to VV$ decays at the leading order (LO), and $\app$ = $\amm$. CP-violating effects in the SM arise beyond the LO but are at the level of $10^{-5}$~\cite{Gritsan:2022php}. Therefore, CP conservation is quite a mild assumption that greatly reduces the statistical uncertainty in the determination of the amplitudes. Because $\app = \amm$, the moduli of the three parameters can be determined from either measurements of $A_{20}^1$ or $A_{20}^2$, and the two statistically-independent measurements can be combined.
The relative sign between $\app$ and $\azz$ is fixed by the Lorentz structure of the vertex~\cite{Aguilar-Saavedra:2022wam}.

 \section{Sensitivity in $H \to ZZ$}

In this section we assess the statistical uncertainty in the measurement of various entanglement observables in $pp \to H \to ZZ \to 4 \ell$  the LHC, using Run 2 + Run 3 data, and at the HL-LHC. For the calculation of the expected number of events we use state-of-the art values of the Higgs production cross section and branching ratio into four electrons or muons. The cross section at next-to-next-to-next-to-leading order is 48.61 pb, 52.23 pb and 54.67 pb at centre-of-mass energies of 13, 13.6 and 14 TeV~\cite{Cepeda:2019klc}, and the Higgs branching ratio decay into four leptons (electrons or muons) is $1.24 \times 10^{-4}$~\cite{Cepeda:2019klc}. The assumed luminosities are 350 fb$^{-1}$ for Runs 2+3 and 3 ab$^{-1}$ for HL-LHC. In order to have a more realistic estimate of the number of events in each case a lepton detection efficiency of 0.7 is assumed, yielding an overall detection efficiency of 0.25. This efficiency accounts for the minimum transverse momentum ($p_T$) thresholds required for lepton detection. We do not include any trigger requirement. The presence of four leptons from the Higgs decay, some of them with significant $p_T$, is expected to fulfill one or many of the trigger conditions for one, two, or three leptons~\cite{trigger}. In addition, we include the efficiency of the invariant mass cut required to remove the interference in same-flavour final states (see the appendix). Overall, the expected number of events for Runs 2+3 and HL-LHC are $N = 490$ and $N = 4500$, respectively.

We do not include backgrounds in our analysis. The $H \to ZZ \to 4\ell$ signal is quite clean, and its main background is the electroweak process $pp \to ZZ/Z\gamma \to 4\ell$, which is about 4 times smaller at the Higgs peak~\cite{CMS:2023gjz}. Although a background subtraction is necessary to obtain the relevant signal distributions, the main effect of the presence of this small background is a slight increase in the statistical uncertainty of the measurement.

The statistical uncertainty is estimated by performing pseudo-experiments. In each pseudo-experiment, a subset of $N$ random events is drawn from the total event set, and for this subset the density operators are calculated as discussed in section~\ref{sec:3}, using the parameter determination in the CP-conserving case. Because the number of events is not large, we use three $m_{V^*}$ bins of 20 GeV. From the density operators, the eigenvalues of the principal eigenvectors, the entanglement measures $N$, and $I_3$, signalling violation of the Bell inequalities are obtained. A large number of $2 \times 10^4$ pseudo-experiments is performed in order to obtain the probability density function (p.d.f.) of these quantities.
The coarse binning used is sufficient even for the statistics of the HL-LHC, as it can be checked by comparing the central values obtained from the pseudo-experiments with those previously obtained in section~\ref{sec:3}. A first test is provided by the eigenvalues of the principal eigenvectors. These are collected in Table~\ref{tab:res1}. As it can be readily observed, the agreement is excellent.

\begin{table}[t]
\begin{center}
\begin{tabular}{lccc}
& theory & Runs 2+3 & HL-LHC \\
$\rho_{LS_1}$,  $\lambda_\text{max}$          & $0.966$ & 0.970  & 0.976 \\
$\rho_{LS_{1,2}}$, $\lambda_\text{max}$    & $0.322$ & 0.323 & 0.325 \\
$\rho_{S_1 S_2}$, $\lambda_\text{max}$            & $0.896$ & 0.897 & 0.905 \\
\end{tabular}
\caption{Theoretical predictions for the eigenvalues of the principal eigenvectors, and central values obtained from pseudo-experiments.}
\label{tab:res1}
\end{center}
\end{table}

\begin{table}[t]
\begin{center}
\begin{tabular}{lccccc}
& theory & \multicolumn{2}{c}{Runs 2+3} & \multicolumn{2}{c}{HL-LHC} \\
$L$-$(S_1 S_2)$, $N$ & $0.757$ & $0.75 \pm 0.14$     & $5.3\sigma$    & $0.75 \pm 0.05$   & $\gg 5\sigma$ \\
$S_i$-$(L S_j)$, $N$    & $0.998$ & $1.0 \pm 0.004$ & $\gg 5\sigma$ & $0.998 \pm 0.001$ & $\gg 5\sigma$ \\
$L$-$S_{1,2}$, $N$      & $0.105$ & $0.106 \pm 0.032^*$ & $4.3\sigma$    & $0.102 \pm 0.010$ & $\gg 5\sigma$ \\
$S_1$-$S_2$, $N$       & $0.843$ & $0.85 \pm 0.05^*$   & $\gg 5\sigma$  & $0.857 \pm 0.014$ & $\gg 5\sigma$ \\
$S_1$-$S_2$, $\mathcal{O}_\text{Bell}$ & $2.507$  &  $2.51 \pm 0.11^*$   & $3.8\sigma$   & $2.539 \pm 0.033$ & $\gg 5\sigma$ \\
$S_1$-$S_2$, $\mathcal{O}'_\text{Bell}$ (helicity)
                          & $2.691$  &  $2.65 \pm 0.09^*$   & $5.7 \sigma$  & $2.678 \pm 0.035$ & $\gg 5\sigma$ \\
\end{tabular}
\caption{Theoretical predictions of several entanglement observables, and their central values and statistical uncertainties obtained from pseudo-experiments. For those marked with an asterisk, the p.d.f. is not Gaussian but skew-normal (see the text). The statistical significance for $N > 0$, or $I_3 > 2$, is also included.}
\label{tab:res2}
\end{center}
\end{table}

\begin{figure}[p]
\begin{center}
\begin{tabular}{cc}
\includegraphics[height=5.5cm,clip=]{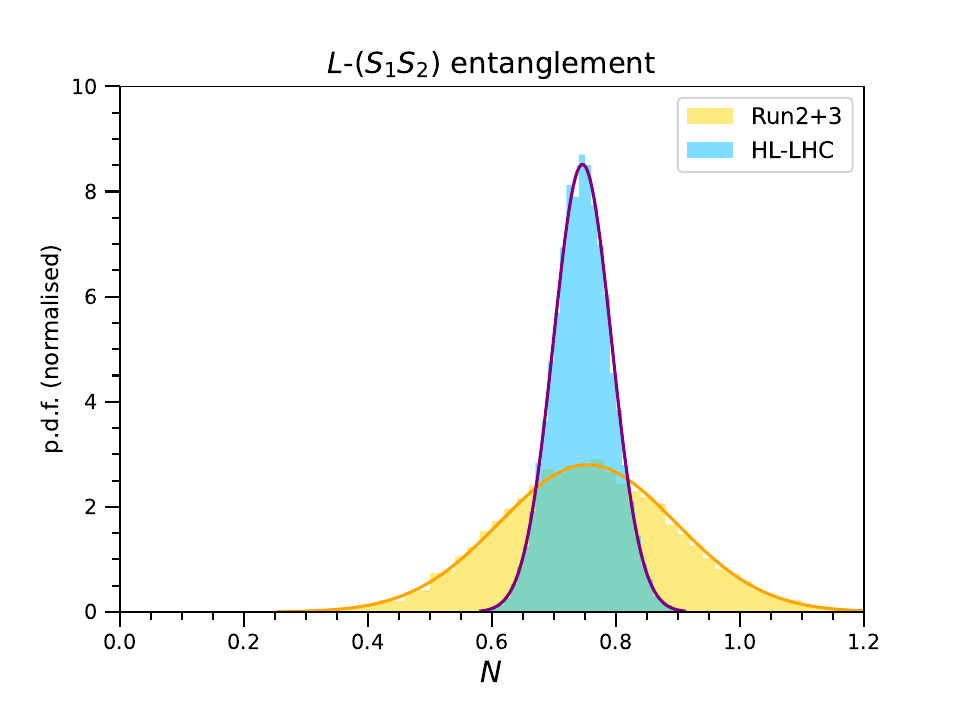} &
\includegraphics[height=5.5cm,clip=]{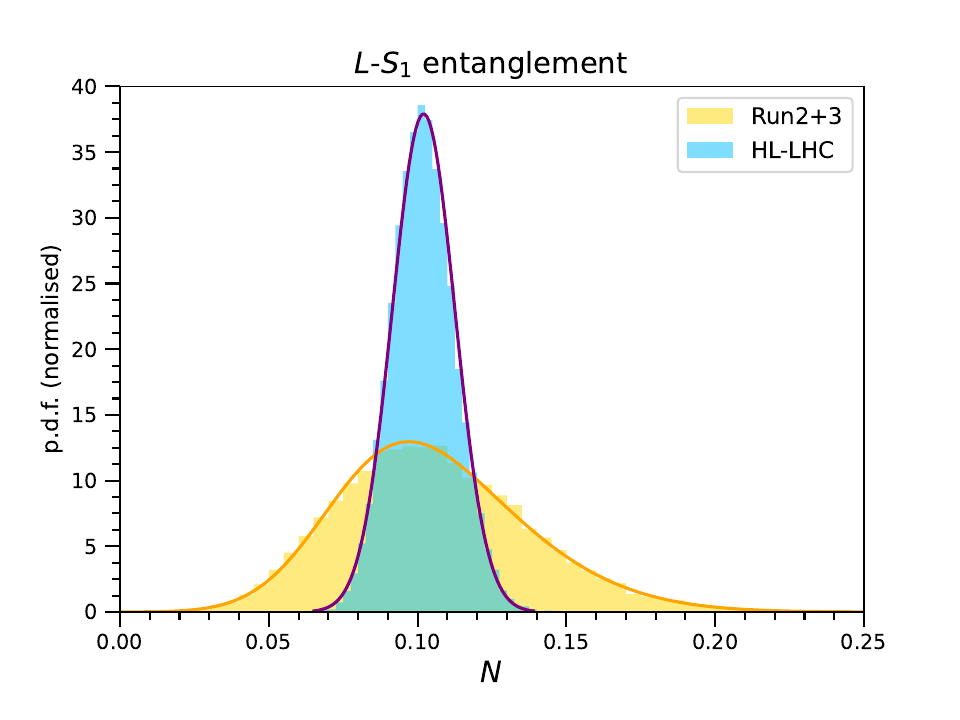}  \\
\includegraphics[height=5.5cm,clip=]{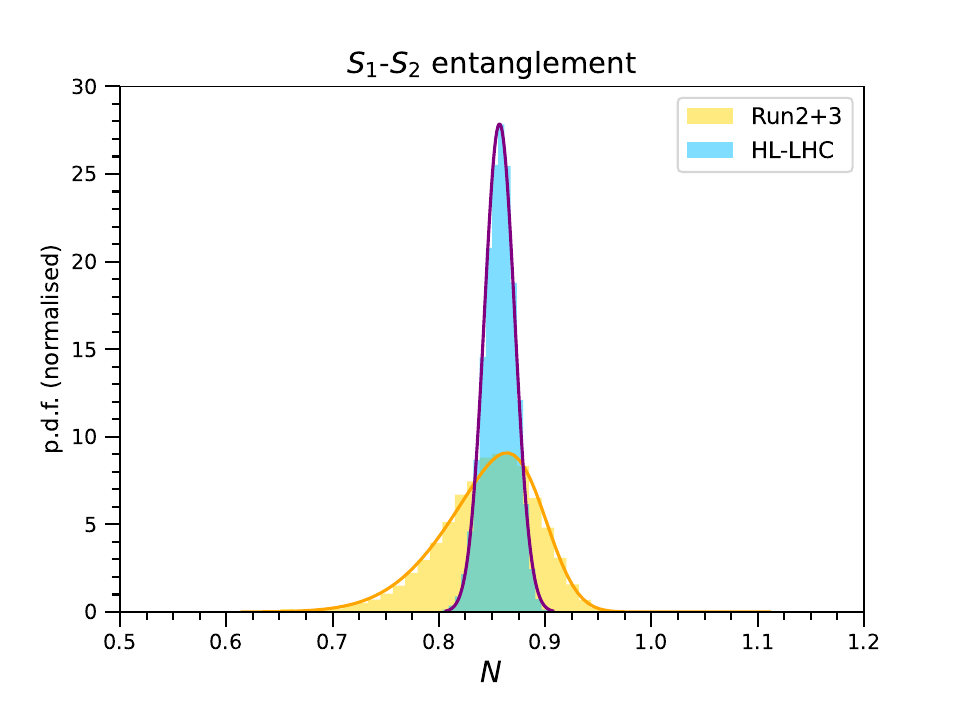} &
\includegraphics[height=5.5cm,clip=]{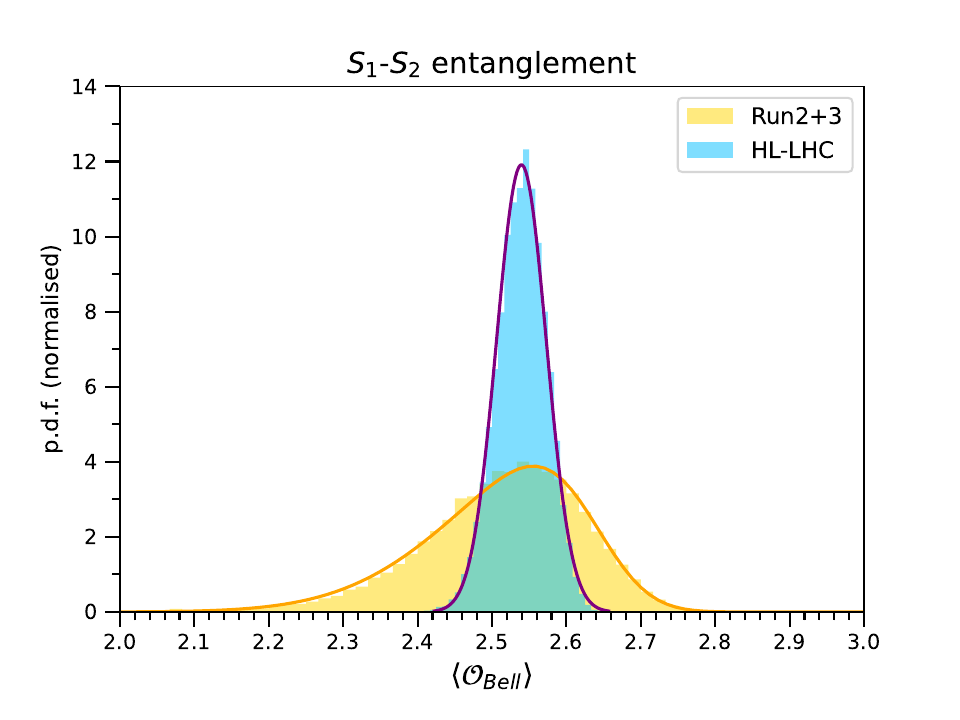} 
\end{tabular}
\caption{Probability density functions for entanglement observables in the canonical basis, as obtained from the pseudo-experiments. The solid lines represent the best-fit Gaussian or skew-normal distributions.}
\label{fig:pdf}
\end{center}
\end{figure}

\begin{figure}[p]
\begin{center}
\includegraphics[height=5.5cm,clip=]{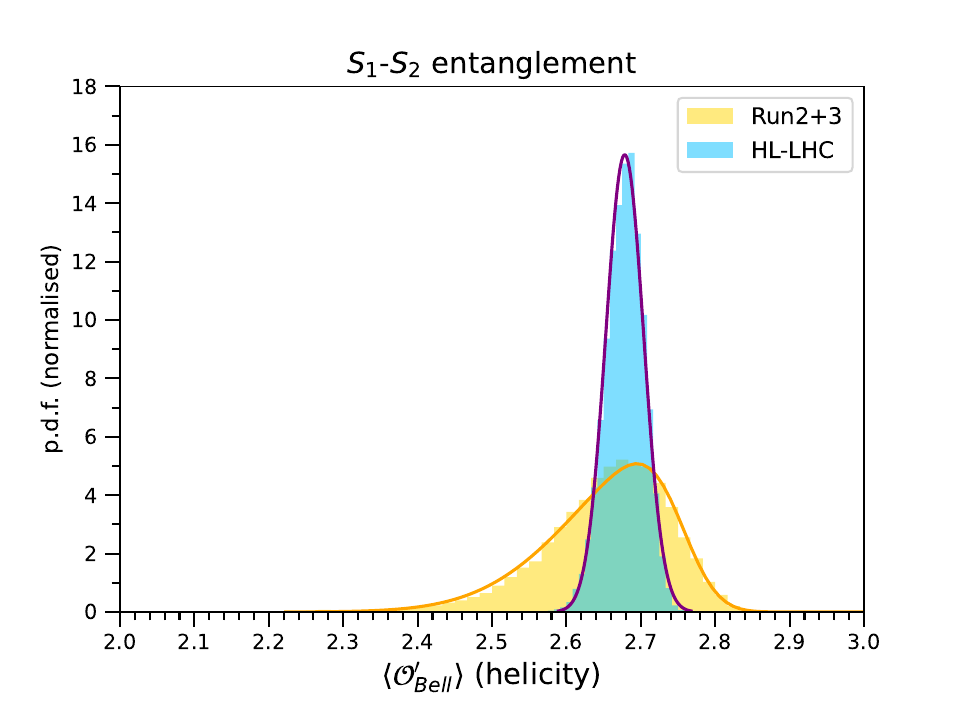} 
\caption{Probability density functions for $\langle \mathcal{O}'_\text{Bell}\rangle$ in the helicity basis, as obtained from the pseudo-experiments. The solid lines represent the best-fit Gaussian or skew-normal distributions.}
\label{fig:pdf2}
\end{center}
\end{figure}

The central values and statistical uncertainties for the entanglement observables of interest are presented in Table~\ref{tab:res2}, together with their theoretical prediction from section~\ref{sec:3}.\footnote{For $\mathcal{O}'_\text{Bell}$ in the helicity basis the same binned method of section~\ref{sec:3} is employed, but using instead Eqs.~(\ref{ec:C}) for the determination of the density operator.}  Again, the agreement between the central values and the theoretical calculations using the full event sample and 2 GeV bins is very good, and sufficient for the statistical uncertainties present. The p.d.f.'s for entanglement observables  are shown in Figs.~\ref{fig:pdf} and \ref{fig:pdf2}. We omit those for $S_i$-$(LS_j)$ and $L$-$S_2$ entanglement; for the former, all the pseudo-experiments give values in an extremely narrow range, and the latter is similar to $L$-$S_1$. For several observables, marked with an asterisk in Table~\ref{tab:res2}, the p.d.f. is not Gaussian for the small number of events expected in LHC Runs 2+3. Still, the p.d.f. is very well approximated by a skew-normal distribution. In those cases, the statistical significances are computed by using the skew-normal distribution with the parameters that best fit the numerical p.d.f. resulting the pseudo-experiments. The statistical significances for entanglement, $N > 0$, and violation of Bell inequalities, $I_3 > 2$, are also presented in Table~\ref{tab:res2}. 

The improvement of statistical uncertainties achieved with this binned reconstruction method, with respect to
previous results for $S_1$-$S_2$ entanglement~\cite{Aguilar-Saavedra:2022wam}, is remarkable. If systematic uncertainties are under control, it allows to establish genuine tripartite entanglement in $H \to ZZ$ decays beyond the $5\sigma$ level at the current LHC run. Also, it allows to verify the violation of Bell inequalities between the two boson spins beyond the $5\sigma$ level.

\section{Discussion}

This work extends previous entanglement studies in $H \to VV$ in two aspects. First, we have generalised the framework to include $L$, thereby being able to study its entanglement with the weak boson spins. Second, we have devised a method to improve the statistical precision in the determination of density operators from data. This method, necessary to obtain the tripartite density operator involving $L$, involves binning the $m_{V^*}$ distribution and using theoretical input to obtain the density operator $\rho_{L S_1 S_2}^{(k)}$ for each bin, from measurements of robust observables. The density operator $\rho_{L S_1 S_2}$ for the full decay phase space is then obtained with a weighted sum of the different bin contributions.
This determination is especially resilient against statistical fluctuations in data when CP conservation is assumed in the decay. Notably, it can also be used to improve the extraction of $S_1 S_2$ operators in the helicity basis, considered in previous work. 

While we have focused on the density operators integrated on the full $m_{V^*}$ range for simplicity, experimental measurements in bins are possible provided there are sufficient statistics. A potential problem for our approach would be event migration between bins, but we expect the impact of this effect in our results to be limited, due to the great similarity between the predictions obtained for 20 GeV bins and 2 GeV bins.

We have not addressed detailed sensitivity estimations for $H \to WW$. Still, a few remarks are in place. In the semi-leptonic channel $WW \to \ell \nu q \bar q'$, Ref.~\cite{Fabbri:2023ncz} used charm tagging to identify the jet corresponding to an initial down-type quark, to be used as spin analyser. Within our framework, where we reconstruct the density operator from measurements of $A_{20}$, the quark identification is not necessary because $Y_2^0 (\theta,\phi) = Y_2^0(\pi-\theta,\phi+\pi)$. Similarly, the yet unexplored decay channel $H \to ZZ \to \ell^+ \ell^- q \bar q$ offers good prospects for experimental measurements.

Finally, our estimation of the expected statistical uncertainties for $H \to ZZ \to 4\ell$ offers excellent prospects, improving over previous results for $S_1$-$S_2$ entanglement~\cite{Aguilar-Saavedra:2022wam} that also assumed CP conservation in the decay. For most entanglement observables considered, including violation of Bell inequalities in the $S_1 S_2$ system, a statistical significance over $5 \sigma$ is expected for the combination of LHC Run 2 and Run 3 data, and for all observables the statistical significance is way above the $5\sigma$ level at HL-LHC.

\section*{Acknowledgements}
This work has been supported by the Spanish Research Agency (Agencia Estatal de Investigaci\'on) through projects PID2019-110058GB-C21, PID2022-142545NB-C21 and CEX2020-001007-S funded by MCIN/AEI/10.13039/501100011033, and by Funda\c{c}{\~a}o para a Ci{\^e}ncia e a Tecnologia (FCT, Portugal) through the project CERN/FIS-PAR/0019/2021.

\appendix
\section{Identical particles in $H \to ZZ$ decay}

The decays $H \to ZZ \to 4e$ and $H \to ZZ \to 4\mu$ involve identical particles in the final state, and two contributing diagrams where they are interchanged. While one cannot properly speak about which $Z$ boson decayed to which opposite-sign pair, because of the interference between diagrams, by a suitable invariant mass selection one can reduce that interference to negligible levels.

We test the effect of identical-particle exchange by generating a sample of $pp \to H \to 4e$ with $7.8 \times 10^6$ events. There are two possible pairings of opposite-sign leptons, `a' and `b', and for each combination we have two invariant masses $m_{e_1^+ e_1^-}$, $m_{e_2^+ e_2^-}$. We select the pairing that produces an opposite-sign pair with the maximum invariant mass, namely the maximum among $m_{e_1^+ e_1^-}^\text{(a)}$, $m_{e_2^+ e_2^-}^\text{(a)}$, $m_{e_1^+ e_1^-}^\text{(b)}$, $m_{e_2^+ e_2^-}^\text{(b)}$. Once the lepton pairing is selected, we identify $V_1$ as the $Z$ boson decaying into the highest invariant mass pair, and measure angular distributions as described in section~\ref{sec:5}. We apply the selection $m_{V_1} \geq 80$ GeV to reduce the interference effects. This cut has an efficiency of 0.77.

The effect in angular distributions of identical-particle exchange can be assessed by comparing the values of non-zero coefficients in the four-dimensional distribution (\ref{ec:dist4D}), for the $H \to ZZ \to e^+ e^- \mu^+ \mu^-$ and the $H \to ZZ \to 4e$ samples. We collect them in the left panel of Table~\ref{tab:idpart}.
In addition, one can compare the principal eigenvalues $\lambda_\text{max}$ of the different density operators, extracted from data using the model-independent method outlined in section~\ref{sec:5}, and  entanglement measures $N$ between different subsystems. We collect them in the right panel of Table~\ref{tab:idpart}. It turns out that the differences have not practical importance even for the higher statistics collected at the HL-LHC.

\begin{table}[htb]
\begin{center}
\begin{tabular}{lcc}
& $e^+e^-\mu^+\mu^-$ & $4e$ \\ 
$B_2 A_{20}^1$  & $-0.636$ & $-0.639$ \\
$B_2 A_{20}^2$ & $-0.634$ & $-0.640$ \\
$B_1^2 C_{111\smo}$ & $0.10$ & $0.21$ \\
$B_1^2 C_{1010}$ & $-0.067$ & $-0.11$ \\
$B_2^2 C_{222\smt}$ & 0.76 & 0.72 \\
$B_2^2 C_{212\smo}$ & $-1.20$ & $-1.16$ \\
$B_2^2 C_{2020}$ & 1.76 & 1.71
\end{tabular}
\quad \quad 
\begin{tabular}{lcc}
& $e^+e^-\mu^+\mu^-$ & $4e$ \\ 
$\rho_{LS_1 S_2}$, $\lambda_\text{max}$    & $0.966$ & $0.961$ \\
$\rho_{LS_i}$, $\lambda_\text{max}$   & $0.322$ & $0.321$ \\
$\rho_{S_1 S_2}$, $\lambda_\text{max}$ & $0.896$ & $0.890$ \\
$S_i$-$(L S_j)$, $N$  & $0.998$ & $1.0$ \\
$L$-$(S_1 S_2)$, $N$ & $0.757$ & $0.776$ \\
$L$-$S_i$, $N$       & $0.105$ & $0.108$ \\
$S_1$-$S_2$, $N$     & $0.843$ & $0.835$
\end{tabular}
\caption{Comparison of different quantities between the decays $H \to ZZ \to e^+ e^- \mu^+ \mu^-$ and $H \to ZZ \to 4e$ (see the text).}
\label{tab:idpart}
\end{center}
\end{table}

\end{document}